\pdfoutput=1
\RequirePackage{ifpdf}
\ifpdf % We are running pdfTeX in pdf mode
\documentclass[pdftex]{sigma}
\else
\documentclass{sigma}
\fi

\usepackage{mathrsfs}

\newcommand{\so}{\scriptscriptstyle \rm I}
\newcommand{\st}{\scriptscriptstyle \rm I\hspace{-1pt}I}

\newcommand\cH{{\mathcal H}}

\newcommand{\kab}{\tilde \kappa}
\newcommand{\ka}{ \kappa}
\newcommand{\kp}{ \kappa^+}
\newcommand{\km}{ \kappa^-}

\newcommand\CC{\mathbb C}

\def\<{\langle}
\def\>{\rangle}
\def\rvec{|0\>}
\def\lvec{\<0|}
\def\bbb{\mathbb{B}}
\def\ccc{\mathbb{C}}

\begin{document}

\allowdisplaybreaks

\newcommand{\arXivNumber}{1506.06550}

\renewcommand{\PaperNumber}{099}

\FirstPageHeading

\ShortArticleName{Slavnov and Gaudin--Korepin Formulas for Models without ${\rm U}(1)$ Symmetry}

\ArticleName{Slavnov and Gaudin--Korepin Formulas for Models\\
 without $\boldsymbol{{\rm U}(1)}$ Symmetry: the Twisted XXX Chain}

\Author{Samuel {BELLIARD}~$^\dag$ and Rodrigo A.~{PIMENTA}~$^{\ddag\S}$}

\AuthorNameForHeading{S.~Belliard and R.A.~Pimenta}

\Address{$^\dag$~Laboratoire de Physique Th\'eorique et Mod\'elisation (CNRS UMR 8089),\\
\hphantom{$^\dag$}~Universit\'e de Cergy-Pontoise, F-95302 Cergy-Pontoise, France}
\EmailD{\href{mailto:samuel.belliard@u-cergy.fr}{samuel.belliard@u-cergy.fr}}

\Address{$^\ddag$~Departamento de F\'{\i}sica, Universidade Federal de S\~ao Carlos,\\
\hphantom{$^\ddag$}~Caixa Postal 676, CEP 13565-905, S\~ao Carlos, Brasil}
\EmailD{\href{mailto:pimenta@df.ufscar.br}{pimenta@df.ufscar.br}}

\Address{$^\S$~Physics Department, University of Miami, P.O. Box 248046, FL 33124, Coral Gables, USA}

\ArticleDates{Received September 02, 2015, in f\/inal form December 02, 2015; Published online December 04, 2015}

\Abstract{We consider the XXX spin-$\frac{1}{2}$ Heisenberg chain on the circle with an arbitrary twist. We characterize its spectral problem using the modif\/ied algebraic Bethe anstaz and study the scalar product between the Bethe vector and its dual. We obtain modif\/ied Slavnov and Gaudin--Korepin formulas for the model. Thus we provide a f\/irst example of such formulas for quantum integrable models without ${\rm U}(1)$ symmetry  characterized by an inhomogenous Baxter T-Q equation.}


\Classification{82B23; 81R12}

The study of quantum integrable models with ${\rm U}(1)$ symmetry by the Bethe ansatz (BA) me\-thods \cite{Bax72, Bet31,FST79} provides exact solutions which found applications in a wide range of domains such as: statistical physics, condensed matter physics, high energy physics, mathematical physics, etc. One of the major accomplishments of the method  has been the obtaining of form factors, for models related to~${\mathfrak{gl}}_2$ and~${\mathfrak{gl}}_3$ families of symmetry, in the compact form of determinants~\cite{BPRS12a, BPRS12b,KKMNST, KKMNST2,KMT99, BogIK93L,BPRS12c}. In particular, for models related to the~${\mathfrak{gl}}_2$ symmetry, the key results are the Slavnov~\cite{Sla89} and the Gaudin--Korepin~\cite{Gau,GauMcCoyWu,K} formulas, which provide, respectively, the scalar product between an eigenstate and an arbitrary state and the norm of the eigenstates.

In the case of models without ${\rm U}(1)$ symmetry, the usual BA techniques in general fail to provide a complete description of the spectrum\footnote{In some cases, \looseness=-1 some gauge transformation can allow to apply the ABA, see for example the XYZ spin chain~\cite{Bax72, FTXYZ}. For the XXX case, that we consider here, the ${\rm GL}(2)$ symmetry allows one to restore the~${\rm U}(1)$ symmetry~\cite{deV} and the usual ABA applies,
provided that the twist is a non-singular matrix. Also, in the context of open XXZ spin chains, constraints on the parameters of the model allow one to apply the usual techniques, see~\cite{TQ}.}.
Thus alternative methods have been de\-ve\-loped, for instance, the separation of variables (SoV) \cite{Der2,Der, KMNT15,NT15,Skly92},
 the commuting transfer matrices method~\cite{BBOY}, the functional method \cite{Gal08} or the $q$-Onsager approach~\cite{BK}. Recently, key steps have been accomplished for the Bethe ansatz solution of such models. On the one hand, a new family of inhomogeneous Baxter T-Q equation to
determine the eigenvalues has been proposed by the of\/f-diagonal Bethe ansatz (ODBA) \cite{CYSW13a, CYSW13b}. On the other hand, the construction of the of\/f-shell Bethe vector has been done in the context of a modif\/ied algebraic Bethe ansatz (MABA) approach \cite{Bel14-3,Bel14, BC13,Bel14-2}. Let us remember that previous developments in the BA technique, in particular the obtainment of the eigenvectors of the XXX chain on the segment with upper-triangular boundaries~\cite{BCR12}, brought important insights to the MABA.

Here, we consider the question of the scalar product between the Bethe vectors obtained from the
modif\/ied algebraic Bethe ansatz. For simplicity, we consider the case of the isotropic spin$-\frac{1}{2}$ Heisenberg chain on the circle with an arbitrary twist.
This is the simplest model which can be considered by the MABA. In fact, its solution contains the main features entailed by the method: the spectrum is characterized
by an inhomogeneous Baxter T-Q relation and the of\/f-shell Bethe vector is generated by a modif\/ied creation operator.
In this context, we obtain a~modif\/ied Slavnov formula for the scalar product between an on-shell Bethe state and its of\/f-shell dual. The formula (see~(\ref{slavfor}))
is given in terms of a determinant depending on the Jacobian of the inhomogenous eigenvalue expression. Moreover, it contains a new factor
related to a certain expansion of the Bethe vector. The square of the norm, i.e., the modif\/ied Gaudin--Korepin formula (see~(\ref{MG})), is obtained by a limit. As expected, the case with general integrable open boundary, which also breaks the ${\rm U}(1)$ symmetry, has the same structure and will be presented in a separated publication.

The isotropic spin$-\frac{1}{2}$ Heisenberg chain on the circle with an arbitrary twist is
given by the Hamiltonian
\begin{gather}\label{HXXX}
H=\,\sum_{k=1}^N\big(\sigma^{x}_k\otimes\sigma^{x}_{k+1}+\sigma^{y}_k\otimes\sigma^{y}_{k+1}+\sigma^{z}_k\otimes\sigma^{z}_{k+1}\big)
\end{gather}
subject to the following boundary conditions
\begin{gather*}
 %\label{cs1}
 \gamma \sigma^{x}_{N+1}=
 \frac{ \kab^2+ \kappa^2-\kappa_+^2-\kappa_-^2}{2}\sigma^{x}_1
 +i\frac{\kappa^2-\kab^2-\kappa_+^2+\kappa_-^2}{2}\sigma^{y}_1
 +(\kappa \kappa_- -\kab \kappa_+)\sigma^{z}_1, \\
 %\label{cs2}
 \gamma \sigma^{y}_{N+1}= i\frac{\kab^2- \kappa^2-\kappa_+^2+\kappa_-^2}{2}\sigma^{x}_1
 +\frac{\kab^2+\kappa^2+\kappa_+^2+\kappa_-^2}{2}\sigma^{y}_1
 -i(\tilde \kappa \kappa_++\kappa \kappa_-)\sigma^{z}_1, \\
 %\label{cs3}
  \gamma \sigma^{z}_{N+1}=
(\kappa \kappa_+-\tilde \kappa \kappa_-)\sigma^{x}_1
  +i(\tilde \kappa \kappa_-+\kappa \kappa_+)\sigma^{y}_1
  +(\tilde \kappa \kappa+\kappa_+ \kappa_-)\sigma^{z}_1.
\end{gather*}
The twist parameters  $\{\kappa,\tilde \kappa,\kappa_+,\kappa_-\} \in \CC^4$ are generic and  $\gamma =\kab\kappa -\kappa_+\kappa_-$. The Pauli matri\-ces\footnote{$
\sigma^{z}=\left(\begin{matrix}
       1 & 0\\
      0 & -1      \end{matrix}\right)$,
$\sigma^{+}=\left(\begin{matrix}
       0 & 1\\
      0 & 0      \end{matrix}\right)$,
$\sigma^{-}=\left(\begin{matrix}
       0 & 0\\
      1 & 0      \end{matrix}\right)$,
      $\sigma^{x}=\sigma^{+}+\sigma^{-}$,  $\sigma^{y}=i(\sigma^{-}-\sigma^{+})$.}~$\sigma^{\alpha}_k$ with $\alpha=x,y,z$  act non trivially on the $k$th space of the quantum space $\cH= \otimes_{k=1}^N V_k$ with $V_k=\CC^2$.

The Hamiltonian~(\ref{HXXX}) is integrable and can be considered within the quantum inverse scat\-te\-ring method~\cite{FT84}. Let us brief\/ly recall this formalism.
The key object is the rational $R$-matrix
\begin{gather*}%\label{R-mat}
 R(u)=\frac{u}{c} + P,
\end{gather*}
which acts on $\CC^2\otimes\CC^2$, with $P=\sigma^+\otimes\sigma^-+\sigma^-\otimes\sigma^++\frac{1}{2}(1+ \sigma^z\otimes\sigma^z)$ and $c \in \CC^*$.
From the $R$-matrix, we construct the monodromy matrix
\begin{gather*}
T_a(u)=R_{a1}(u-\theta_1)\cdots R_{aN}(u-\theta_N)=\left(\begin{matrix} t_{11}(u)& t_{12}(u)\\ t_{21}(u)&t_{22}(u)\end{matrix}\right)_a,
\end{gather*}
which acts on $\CC^2\otimes \cH$ and
with $\{\bar \theta\}=\{\theta_1,\dots \theta_N\}\in\CC^N$ being the inhomogeneity parameters.
This monodromy matrix satisf\/ies the RTT relation
\begin{gather*}%\label{RTT}
R_{ab}(u-v)T_a(u)T_b(v)=T_b(v)T_a(u)R_{ab}(u-v)
\end{gather*}
that encodes commutation relations between the operators $\{t_{ij}(u)\}$, see Appendix~\ref{appA}.
The transfer matrix, generating function of the conserved quantities of the model, is given by
\begin{gather*}
t(u)=\operatorname{Tr}_a\big(K_aT_a(u)\big)=\kab t_{11}(u)+\ka t_{22}(u)+\kp t_{21}(u)+\km t_{12}(u),
\end{gather*}
with
\begin{gather}\label{trans}
K=\left(\begin{matrix} \tilde \kappa & \kappa^+\\ \kappa^-&\kappa \end{matrix}\right).
\end{gather}
The commutation relation between transfer matrices with dif\/ferent spectral parameters
\begin{gather*}
[t(u),t(v)]=0,
\end{gather*}
follows from the RTT relation and the ${\rm GL}(2)$ invariance of the $R$-matrix
\begin{gather}\label{gl2invR}
[R_{ab}(u-v),K_aK_b]=0.
\end{gather}
The Hamiltonian (\ref{HXXX}) is given by
\begin{gather*}%\label{Hfromt}
H=2c  \frac{d}{d u}\big(\ln (t(u) )\big)\big|_{u\to 0, \, \theta_i\to 0}-N,
\end{gather*}
and thus its spectral problem is the same of the one of the transfer matrix.

The diagonalization of the transfer matrix can be obtained by means of the MABA \cite{Bel14-3, Bel14,BC13,Bel14-2} and leads to an
inhomogeneous Baxter T-Q equation. In order to do that, we introduce the following transformation of the twist matrix~(\ref{trans})
\begin{gather*}%\label{tfmaba1}
K=LDL
\end{gather*}
with
\begin{gather*}%\label{tfmaba2}
L=\mu^{\frac{1}{2}}\left(\begin{matrix} 1 & \frac{\rho}{\kappa^-}\\  \frac{\rho}{\kappa^+}&1 \end{matrix}\right), \qquad
D=\left(\begin{matrix} \kab-\rho &0\\ 0&\kappa-\rho \end{matrix}\right), \qquad \mu=\frac{\kab+\ka-\rho}{\kab+\ka-2\rho}
\end{gather*}
and
\begin{gather*}%\label{contrho}
\rho^2-(\kab+\kappa)\rho+\kp\km=0.
\end{gather*}
By means of this transformation, we can obtain a modif\/ied monodromy matrix $\overline T(u)=L T(u) L$ with entries given by modif\/ied operators $\{\nu_{ij}(u)\}$. They are expressed in terms of the initial $\{t_{ij}(u)\}$ operators by
\begin{gather*}%\label{tfT}
\nu_{11}(u)= \mu\left(t_{11}(u)+\frac{\rho}{\kappa^+} t_{12}(u)+\frac{\rho}{\kappa^-} t_{21}(u)+\frac{\rho^2}{\kappa^+\kappa^-} t_{22}(u)\right),\\
\nu_{12}(u)= \mu \left(t_{12}(u)+\frac{\rho}{\kappa^-}( t_{11}(u)+t_{22}(u))+\left(\frac{\rho}{\kappa^-}\right)^2 t_{21}(u)\right),\\
\nu_{21}(u)= \mu \left(t_{21}(u)+\frac{\rho}{\kappa^+}( t_{11}(u)+t_{22}(u))+\left(\frac{\rho}{\kappa^+}\right)^2 t_{12}(u)\right),\\
\nu_{22}(u)= \mu \left(t_{22}(u)+\frac{\rho}{\kappa^+} t_{12}(u)+\frac{\rho}{\kappa^-} t_{21}(u)+\frac{\rho^2}{\kappa^+\kappa^-} t_{11}(u)\right).
\end{gather*}
It follows that the transfer matrix has a modif\/ied diagonal form given by
\begin{gather*}%\label{tft}
t(u)=\operatorname{Tr}\big(D\bar T(u)\big)=(\kab-\rho)\nu_{11}(u)+(\kappa-\rho)\nu_{22}(u).
\end{gather*}
To construct the Bethe vector we use the usual highest weight representation and the highest weight vector
\begin{gather*}
\rvec= \left( \begin{matrix}1\\ 0\end{matrix} \right)^{\otimes^N}.
\end{gather*}
The actions of the operators $ \{t_{ij}(u)\}$ on it are given by
\begin{gather*}%\label{tvac}
t_{ii}(u)\rvec=\lambda_i(u)\rvec, \qquad t_{21}(u)\rvec=0,
\end{gather*}
with
\begin{gather}\label{lam}
\lambda_1(u)=\prod_{i=1}^N\frac{u-\theta_i+c}{c}, \qquad \lambda_2(u)=\prod_{i=1}^N\frac{u-\theta_i}{c}.
\end{gather}
For the dual Bethe vector we use the dual highest weight vector
\begin{gather*}
\lvec= (1, \, 0 )^{\otimes^N},
\end{gather*}
with the actions
\begin{gather*}%\label{tvac}
\lvec t_{ii}(u)=\lambda_i(u)\lvec, \qquad \lvec t_{12}(u)=0, \qquad \langle 0\rvec=1.
\end{gather*}
For the operators $\{\nu_{ij}(u)\}$, we have a modif\/ied action on the highest weight vector. We can show that
\begin{gather}\label{nu11vac}
 \nu_{11}(u)\rvec=\lambda_1(u)\rvec+\frac
{\rho}{\kp}\nu_{12}(u)\rvec,\\
\label{nu22vac}
 \nu_{22}(u)\rvec=\lambda_2(u)\rvec+\frac
{\rho}{\kp}\nu_{12}(u)\rvec,\\
 \nu_{21}(u)\rvec=\frac
{\rho}{\kp}\big(\lambda_1(u)+\lambda_2(u)\big)\rvec+\left(\frac{\rho}{\kp}\right)^2\nu_{12}(u)\rvec. \label{nu21vac}
\end{gather}
We have thus all the ingredients to implement the MABA. We will use the notation $\bar u $ with $\# \bar u =M$ for the set of $M$ variables $\{u_1,u_2,\dots,u_M\}$. If the element $u_i$ is removed, we note $\bar u_i =\{u_1,u_2,\dots,u_{i-1},u_{i+1},\dots,u_M\}$. If we also remove the element $u_j$, we note $\bar u_{ij}=\bar u/\{u_i,u_j\}$. For products of functions (see~(\ref{basicfuncfg}) bellow) or of operators $\left\{\nu_{ij}(u)\right\}$, we use the convention
\begin{gather*}
 g(u,\bar u)=\prod_{i=1}^Mg(u,u_i), \qquad g(\bar v,\bar u)=\prod_{i=1}^M\prod_{j=1}^Mg(v_j,u_i), \\
 g(u_i,\bar u_i)=\prod_{j=1, j\neq i}^Mg(u_i,u_j),  \qquad \nu_{ij}(\bar u)=\prod_{k=1}^M\nu_{ij}(u_k).
\end{gather*}
The functions
\begin{gather} \label{basicfuncfg}
g(u,v)=\frac{c}{u-v},\qquad f(u,v)=1+g(u,v)=\frac{u-v+c}{u-v}
\end{gather}
will be widely used.

Let us consider the vector
\begin{gather}\label{BV}
\bbb^{M}(\bar u) = \nu_{12}(u_1)\cdots \nu_{12}(u_M)\rvec=\nu_{12}(\bar u)\rvec,
\end{gather}
and act with the transfer matrix on it. In order to perform this calculation we need to f\/ind the action of the operators $\{\nu_{ij}(u)\}$ on the vector~(\ref{BV}).
Using the ${\rm GL}(2)$ invariance~(\ref{gl2invR}), it is easy to see that the new operators satisfy the same commutation relations~(\ref{comsl21}), (\ref{comsl22}), (\ref{comsl23}) of the operators $\{t_{ij}(u)\}$, see Appendix~\ref{appA}. Thus, using these commutation relations and the action (\ref{nu11vac}), (\ref{nu22vac}), (\ref{nu21vac}), we can show that
\begin{gather}
 \nu_{12}(u)\bbb^{M}(\bar u)=\bbb^{M+1}(u,\bar u),\nonumber\\
 \nu_{11}(u)\bbb^{M}(\bar u)=\frac
{\rho}{\kp}\bbb^{M+1}(u,\bar u)+\lambda_1(u)f(\bar u,u)\bbb^{M}(\bar u)\nonumber\\
\hphantom{\nu_{11}(u)\bbb^{M}(\bar u)=}{}  +\sum_{i=1}^Mg(u,u_i)\lambda_1(u_i) f(\bar u_i,u_i)\bbb^{M}(u,\bar u_i),\nonumber\\%\label{nu11BV}
 \nu_{22}(u)\bbb^{M}(\bar u)=\frac{\rho}{\kp}\bbb^{M+1}(u,\bar u)+\lambda_2(u)f(u,\bar u)\bbb^{M}(\bar u)\nonumber\\
\hphantom{\nu_{22}(u)\bbb^{M}(\bar u)=}{}
 +\sum_{i=1}^Mg(u_i,u)\lambda_2(u_i) f(u_i,\bar u_i)\bbb^{M}(u,\bar u_i),\nonumber\\%\label{nu22BV}
 \nu_{21}(u)\bbb^{M}(\bar u)=\left(\frac{\rho}{\kp}\right)^2\bbb^{M+1}(u,\bar u)\label{nu21BV}\\
 \hphantom{\nu_{21}(u)\bbb^{M}(\bar u)=}{}
  +\frac{\rho}{\kp}\left(\Lambda^M_{d}(u,\bar u|1,1)\bbb^{M}(\bar u)+\sum_{i=1}^Mg(u_i,u)E^M_{d}(u_i,\bar u_i|1,1)\bbb^{M}(u,\bar u_i)\right)\nonumber\\
\hphantom{\nu_{21}(u)\bbb^{M}(\bar u)=}{}
+\left(\sum_{i=1}^MF(u,u_i,\bar u_i)\bbb^{M-1}(\bar u_i)+\sum_{1\leq i<j \leq M}G(u,u_i,u_j,\bar u_{ij})\bbb^{M-1}(u,\bar u_{ij})\right),\nonumber
\end{gather}
with
\begin{gather*}
 \Lambda^M_{d}(u,\bar u|x,y)=x f(\bar u,u)\lambda_1(u)+y f(u,\bar u)\lambda_2(u),\\
 E^M_{d}(u_i,\bar u_i|x,y)=-x f(\bar u_i,u_i)\lambda_1(u_i)
+y f(u_i,\bar u_i)\lambda_2(u_i),\\
 F(u,u_i,\bar u_i)=g(u,u_i)\lambda_1(u)\lambda_2(u_i)f(u,\bar u_i)f(\bar u_i,u_i)
  +g(u_i,u)\lambda_1(u_i)\lambda_2(u)f(u_i,\bar u_i)f(\bar u_i,u), \\
 G(u,u_i,u_j,\bar u_{ij})=g(u,u_i)g(u_j,u)\lambda_1(u_i)\lambda_2(u_j)f(u_i,u_j)f(u_i,\bar u_{ij})f(\bar u_{ij},u_j)\\
\hphantom{G(u,u_i,u_j,\bar u_{ij})=}{}
+g(u,u_j)g(u_i,u)\lambda_1(u_j)\lambda_2(u_i)f(u_j,u_i)f(u_j,\bar u_{ij})f(\bar u_{ij},u_i),
\end{gather*}
and where we have used the functional identities
\begin{gather*}%\label{relSS2}
 f(\bar u,u)+\sum_{i=1}^Mg(u,u_i)f(\bar u_i,u_i)=1,\qquad f(u,\bar u)+\sum_{i=1}^Mg(u_i,u)f(u_i,\bar u_i)=1.
\end{gather*}
It follows that the action of the transfer matrix is given by
\begin{gather}
 t(u)\bbb^{M}(\bar u)=\frac{\km}{\mu} \bbb^{M+1}(u, \bar u)+\Lambda^M_{d}(u,\bar u|\kab-\rho,\kappa-\rho)\bbb^{M}(\bar u)\nonumber\\
 \hphantom{t(u)\bbb^{M}(\bar u)=}{}  +\sum_{i=1}^Mg(u_i,u)E^M_{d}(u_i,\bar u_i|\kab-\rho,\kappa-\rho)\bbb^{M}(u,\bar u_i),\label{actiontd}
\end{gather}
where we have used the relation $\frac{\rho}{\kp} (\kab+\ka-2\rho) =\frac{\km}{\mu}$.
The new term $\bbb^{M+1}(u,\bar u)$ is characteristic for models which break
the ${\rm U}(1)$ symmetry.  From the formula~(\ref{actiontd}) we can obtain, by limit the upper triangular case $\km=0$ and the diagonal case $ \km=\kp=0$. The Bethe ansatz solution is then obtained by requiring that  $E^M_{d}(u_i,\bar u_i|\kab-\rho,\kappa-\rho)=0$ for $M=0,\dots,N$. Only the part with positive total spin contributes to the solution.
Here to complete the MABA we must thus f\/ind the action of the operator $\nu_{12}(u)$ on the Bethe vector (\ref{BV}) when $M=N$.
In this case the $2^N$ independent diagonal Bethe vectors with $M\leq N$ and partitions of $N$ variables are contained in the vector, see Appendix~\ref{appB} for the expression of the Bethe vector in terms of the initial opera\-tor~$t_{12}(u)$. This allows us to f\/ind an {\it off-shell} action with a wanted/unwanted form
\begin{gather} \label{cong}
\frac{\km}{\mu} \mathbb{B}^{N+1}(u,\bar u)=\Lambda_g^N(u,\bar u)\mathbb{B}^N(\bar u) +\sum_{i=1}^Ng(u_i,u)E_g^N(u_i,\bar u_i)\mathbb{B}^N(u,\bar u_i),\\
\Lambda_g^N(u,\bar u)=2\rho\lambda_1(u) \lambda_2(u)g(u,\bar u), \qquad E_g^N(u_i,\bar u_i)=2\rho\lambda_1(u_i) \lambda_2(u_i)g(u_i,\bar u_i).\nonumber
\end{gather}
This action can be proved following the method given in~\cite{Cra14} and will be considered elsewhere.
Finally, from~(\ref{actiontd}) and~(\ref{cong}), we obtain the {\it off-shell} equation with generic $\bar u$ with $\# \bar u=N$,
\begin{gather*}%\label{t-B-gen}
t(u)\,\mathbb{B}^N(\bar u) =\Lambda^{N}(u, \bar u)\mathbb{B}^N(\bar u)+
\sum_{i=1}^N g(u_i,u)E^N(u_i,\bar u_i) \mathbb{B}^N(u,\bar u_i),
\end{gather*}
with an {\it inhomogeneous} eigenvalue
\begin{gather}\label{LamL}
\Lambda^{N}(u, \bar u)=(\kab-\rho) \lambda_1(u)f(\bar u,u)+(\ka-\rho)\lambda_2(u)f(u,\bar u)+2\rho\lambda_1(u) \lambda_2(u)g(u,\bar u),
\end{gather}
and an {\it inhomogeneous} Bethe equation
\begin{gather}
E^N(u_i,\bar u_i)=-(\kab-\rho)\lambda_1(u_i) f(\bar u_i,u_i)+(\kappa-\rho)\lambda_2(u_i) f(u_i,\bar u_i)\nonumber\\
\hphantom{E^N(u_i,\bar u_i)=}{} +2\rho\lambda_1(u_i) \lambda_2(u_i)g(u_i,\bar u_i),\label{EBL}
\end{gather}
for $i=1,\dots,N$.

One can proceed in a similar way for the dual Bethe vector
\begin{gather}\label{dualBV}
\ccc^{N}(\bar u)=\lvec \nu_{21}(\bar u),
\end{gather}
and, in particular, obtain
\begin{gather*}
\ccc^{N}(\bar u)t(u)=\Lambda^{N}(u, \bar u)\ccc^N(\bar u)+
\sum_{i=1}^N g(u_i,u)E^N(u_i,\bar u_i) \ccc^N(u,\bar u_i),
\end{gather*}
with the same eigenvalue and Bethe equations of the Bethe vector (\ref{BV}) with $M=N$.

When the Bethe equations are satisf\/ied, i.e., $E^N(u_i,\bar u_i)=0$ for $i=1,\dots, N$, and we consider non-singular solutions of the Bethe equations~\cite{singsol}, the {\it on-shell} Bethe vectors are eigenstates of the transfer matrix
\begin{gather*}
t(u) \mathbb{B}^N(\bar u) =\Lambda^{N}(u, \bar u)\mathbb{B}^N(\bar u), \qquad \ccc^{M}(\bar u)t(u)=\Lambda^{N}(u, \bar u)\ccc^N(\bar u).
\end{gather*}
Let us remark that the completeness of the solution
given by~(\ref{LamL}),~(\ref{EBL})
has been numerically checked for chains with small size. It should be interesting to prove it along the lines of~\cite{MTV}.

We are now in position to consider the scalar products for the
Bethe vector (\ref{BV}) and (\ref{dualBV}), namely,
\begin{gather}\label{Scal}
S^N(\bar u|\bar v)=\ccc^{N}(\bar u)\bbb^{N}(\bar v).
\end{gather}
From the construction given hereafter,
we obtain the modif\/ied Slavnov formula of the twisted XXX spin chain characterized by the inhomogeneous Baxter T-Q equation~(\ref{LamL}).
When
\begin{gather*}
E^N(u_i,\bar u_i)=0,
\end{gather*}
for $i=1,\dots, N$, the scalar product~(\ref{Scal}) has a compact form given by
\begin{gather}\label{slavfor}
\hat S^N(\bar u, \bar v)=c^N\left(\frac{\mu^2}{\kab+\ka-\rho}\right)^NW_0^N(\bar u) \frac{\operatorname{Det}_N\big(\frac{\partial}{\partial u_i}\Lambda^N(v_j,\bar u)\big)}{\operatorname{Det}_N\big(g(v_i,u_j)\big)}
\end{gather}
with
\begin{gather*}
W_0^N(\bar u)=\left( \frac{\km}{\mu\rho}\right)^N\lvec \nu_{12}(\bar u) \rvec,
\end{gather*}
which is given explicitly in Appendix~\ref{appB}.
When
\begin{gather*}
E^N(v_i,\bar v_i)=0,
\end{gather*}
for $i=1,\dots, N$, the scalar product (\ref{Scal}) is given by
\begin{gather}\label{slavfor2}
\tilde S^N(\bar u, \bar v)=c^N\left(\frac{\mu^2}{\kab+\ka-\rho}\right)^NW_0^N(\bar v)\frac{\operatorname{Det}_N\big(\frac{\partial}{\partial v_i}\Lambda^N(u_j,\bar v)\big)}{\operatorname{Det}_N\big(g(u_i,v_j)\big)}.
\end{gather}
We can obtain the square of the norm by imposing the limit $\bar u=\bar v$. Using the well-known formula for the Cauchy determinant
\begin{gather*}
\operatorname{Det}_N\big(g(v_i,u_j)\big)=\frac{g(\bar v,\bar u)}{\prod\limits_{i<j} g(u_i,u_j)g(v_j,v_i)},
\end{gather*}
we obtain
\begin{gather}\label{MG}
{\mathscr N}^N(\bar u)=\left(\frac{\mu^2}{\kab+\ka-\rho}\right)^NW_0^N(\bar u)\left(\prod_{i<j}^Ng(u_i,u_j)g(u_j,u_i)\right)\operatorname{Det}_N (G_{ij} ),
\end{gather}
where the matrix elements $G_{ij}$, for $i,j=1,\dots, N$, are given by
\begin{gather*}
 G_{ii}=2\rho  c  (\lambda_2(u_i) \partial_{u_i}\lambda_1(u_i)+\lambda_1(u_i) \partial_{u_i}\lambda_2(u_i))\\
\hphantom{G_{ii}=}{} +(-1)^N(\kab -\rho)\left(c\, h(\bar u_i,u_{i})\partial_{u_i} \lambda_1(u_i)- \lambda_1(u_i)\sum_{j=1,j\neq i}^Nh(\bar u_{ij},u_i)\right)\\
\hphantom{G_{ii}=}{}
+(\ka -\rho)\left(c  h(u_{i},\bar u_i)\partial_{u_i} \lambda_2(u_i) +\lambda_2(u_i)\sum_{j=1,j\neq i}^Nh(u_i,\bar u_{ij})\right)\\
 G_{ij}=(-1)^N(\kab -\rho)\lambda_1(u_j)h(\bar u_{ij},u_j)-(\ka -\rho)\lambda_2(u_j)h(u_j,\bar u_{ij}) \qquad \text{for} \quad i\neq j,
\end{gather*}
with
\begin{gather*}
h(u,v)=\frac{f(u,v)}{g(u,v)}=\frac{u-v+c}{c},
\end{gather*}
and where we have applied the l'Hospital's rule to f\/ind
\begin{gather} \label{lim1}
G_{ij}=\lim_{v_j\to u_j} c\frac{\frac{\partial}{\partial u_i}\Lambda^N(v_j,\bar u)}{g( v_j, \bar u) }.
\end{gather}

From the conjectured modif\/ied Slavnov and Gaudin--Korepin formulas we remark that the ratio, needed for the calculation of the correlations functions, is independent of the $W_0^N(\bar u)$ and of the constant $\big(\frac{\mu^2}{\kab+\ka-\rho}\big)^N$. Thus the relevant part of the formula for  the correlations functions is given by the Jacobian of the inhomogeneous eigenvalue~(\ref{LamL}) and  it limits~(\ref{lim1}).

To obtain this conjecture we have proceeded in the following way.
We start from the following hypothesis:
\begin{itemize}\itemsep=0pt
\item We can impose the Bethe equations~(\ref{EBL}), $E^N(v_i,\bar v_i)=0
$
for $i=1,\dots, N$, by linearizing the quadratic terms $\lambda_1(v_i)\lambda_2(v_i)$  in terms of  $\lambda_1(v_i)$ and~$\lambda_2(v_i)$.
\item The usual Slavnov formula must be recovered in the~${\rm U}(1)$ symmetric limit and thus the modif\/ied Slavnov formula contain the determinant of the Jacobian of the inhomogenous eigenvalue~(\ref{LamL}).
\end{itemize}
For the ${\rm U}(1)$ symmetric case ($\kp=\km=0$), the scalar product between an on-shell Bethe vector and an of\/f-shell Bethe vector is given by the Slavnov formula \cite{Sla89}. If the $\bar v$ are on-shell, i.e., $E^M_{d}(v_i,\bar v_i|\kab,\kappa)=0$ for $i=1,\dots,M$, we have
\begin{gather*}
\tilde{S}_d^M(\bar u,\bar v)=\left(\frac{c}{\kab}\right)^M \lambda_2(\bar v)\frac{{\operatorname{Det}_M\big(\frac{\partial}{\partial v_i}\Lambda_d^M(u_j,\bar v,\kab,\ka)\big)}}{\operatorname{Det}_M (g(u_i,v_j) )}.
\end{gather*}
From the f\/irst term of action the~(\ref{nu21BV}), for $M=N$ and using~(\ref{cong}), we can extract the leading term of the scalar product with $N$ quadratic terms: $\lambda_1(u_i)\lambda_2(u_i)$ with $i=1,\dots, N$.
If we impose the Bethe equations~(\ref{EBL}), $E^N(v_i,\bar v_i)=0
$
for $i=1,\dots, N$, by linearizing the quadratic terms $\lambda_1(v_i)\lambda_2(v_i)$  in terms of  $\lambda_1(v_i)$ and $\lambda_2(v_i)$,
this leading term is invariant and is the only contribution at the top order~$3N$ in the~$\lambda_i$. This term is proportional to
\begin{gather*}
W_0^N(\bar v)\prod_{i=1}^N\Lambda^N_g(u_{i},\bar v).
\end{gather*}
It allows us to f\/ix the functional to complete the determinant part of the Slavnov formula~(\ref{slavfor2}). Indeed, the functional is the same, up to a constant,  when we consider the leading coef\/f\/icient in the~$\lambda_i$  and use the identity
\begin{gather*}
\operatorname{Det}_N\left(\frac{\partial}{\partial v_i}\Lambda_g^N(u_j,\bar v)\right)=\frac{1}{c^N}  \operatorname{Det}_N\big(g(u_j,v_i)\big)\prod_{i=1}^N\Lambda_g^N(u_i,\bar v).
\end{gather*}
In the ${\rm U}(1)$ symmetry limit, $\rho=0$, we have $ W_0^N(\bar v)=\lambda_2(\bar v)(\frac{\ka}{\kab}+1)^N$ and restore the usual Slavnov formula up to a~constant. The formula~(\ref{slavfor}) can de derived, also up to a constant, in a~similar way starting from the action of the operator~$\nu_{12}(u)$ on the dual Bethe vector~(\ref{dualBV}). This way to f\/ind the modif\/ied Slavnov formula does not f\/ix the constant.

Let us f\/ix this constant by considering the simplest case $N=1$ and discuss the dif\/f\/iculties to f\/ind the good parametrization for the of\/f-shell scalar product, which allows us to f\/ind the modif\/ied Slavnov formula by linearizing the quadratic term of the Bethe equation, for general~$N$. In the case $N=1$, the good one  is
\begin{gather}\label{goodN1}
 S^1(u, v)=\mu\left(S_d^1(u, v)+\frac{\mu}{\kab+\ka-\rho}\big(\Lambda^1_g(u,v)W_0^1(v)+\Lambda^1_g(v,u)W_0^1(u)\big)\right)
\end{gather}
with
\begin{gather*}
W_0^1(u)=\lambda_1(u)+\lambda_2(u), \qquad S_d^1(u, v)=g(u,v)(\lambda_1(v)\lambda_2(u)-\lambda_1(u)\lambda_2(v)).
\end{gather*}
To obtain the Slavnov formula we use the Bethe equation to linearize the quadratic term that depends on~$v$
\begin{gather}\label{linN1}
\lambda_1(v)\lambda_2(v)=\frac{1}{2\rho}\big((\kab-\rho)\lambda_1(v)-(\ka-\rho)\lambda_2(v)\big)
\end{gather}
and thus we arrive to our result
\begin{gather*}
S_{E(v)=0}^1(u, v)=\frac{\mu^2}{\kab+\ka-\rho}g(v,u)W_0^1(v)\big((\kab-\rho)\lambda_1(u)-(\ka-\rho)\lambda_2(u)-2\rho\lambda_1(u)\lambda_2(u)\big).
\end{gather*}
If we also linearize the second quadratic term,
we f\/ind zero for $u\neq v$, which shows the ortho\-go\-nality of the Bethe vectors and
the modif\/ied Gaudin--Korepin formula~(\ref{MG}) for  $u=v$.

The parametrization~(\ref{goodN1}) is not unique; for instance, from the projection on the $\{t_{ij}(u)\}$ operator, see Appendix~\ref{appB}, we f\/ind
\begin{gather*}
S^1(u, v)=\mu^2 S_d^1(u, v)+\mu^2 \frac{\rho^2}{\kp\km} \big(\lambda_1(u)+\lambda_2(u)\big)\big(\lambda_1(v)+\lambda_1(v)\big)
\end{gather*}
that is equivalent to~(\ref{goodN1}) when we specify the explicit form of the~$\lambda_i(u)$~(\ref{lam}). In this case the quadratic term $\lambda_1(v)\lambda_2(v)$ does not appear and the formula~(\ref{linN1}) could not be used directly.
Moreover, the recursion relation~(\ref{nu21BV}) gives also another parametrization in order 3 in the $\lambda_i(u)$ like~(\ref{goodN1}) but with terms of the form $\lambda_i(u)\lambda_i(v)$.
In this case the linearization of the quadratic term~(\ref{linN1}) could be used but does not lead directly to the modif\/ied Slavnov formula.
A systematic way to f\/ix the good parametrization of the of\/f-shell scalar product (i.e., a parametrization that reduces to the modif\/ied Slavnov formula, when we linearize the quadratic terms $\lambda_1(v_i)\lambda_2(v_i)$ from the inhomogenous Bethe equations) remains an open problem for the moment and will be discussed elsewhere. The conjecture can be tested exactly for~$N=1$, which allows us to f\/ix the constant, and then the cases $N=2$ and $N=3$ have been checked numerically to support the conjecture.

Finally, let us point out another problem which is worth to be considered: the construction of an explicit algebraic link between the MABA solution and
the solution obtained by means of the usual ABA, see, e.g.,~\cite{RMG03}.
One f\/irst step to address this problem, at least at the spectral
level, is to equate the eigenvalue expression found from the MABA~(\ref{LamL}) with the ones obtained from the usual ABA. The simplest eigenvalue from the usual ABA is given by
\begin{gather*}
\alpha \lambda_1(u)+(\kappa+\tilde \kappa -\alpha) \lambda_2(u)
\end{gather*}
with $\alpha=\frac{1}{2}(\kappa+\tilde \kappa+\sqrt{(\kappa-\tilde \kappa)^2+4\kappa^+\kappa^-})$ an eigenvalue of the twist matrix~$K$. As a result, we can obtain constraints on the parameters $\bar u$ that, up to symmetrization, provide one solution of the Bethe equations~(\ref{EBL}). More details will be given elsewhere.

\appendix

\section{Rational functions and commutations relations}  \label{appA}
Let us introduce the commutation relations of the $t_{ij}(u)$ given by
\begin{gather*}%\label{com}
[t_{ij}(u),t_{kl}(v)]=g(u,v)\big(t_{kj}(v)t_{il}(u)-t_{kj}(u)t_{il}(v)\big).
\end{gather*}
In particular, we will only use the following ones
\begin{gather}\label{comsl21}
 t_{11}(u)t_{12}(v)=f(v,u)t_{12}(v)t_{11}(u)+g(u,v)t_{12}(u)t_{11}(v),\\
\label{comsl22}
 t_{22}(u)t_{12}(v)=f(u,v)t_{12}(v)t_{22}(u)+g(v,u)t_{12}(u)t_{22}(v),\\
\label{comsl23}
 t_{21}(u)t_{12}(v)=t_{12}(u)t_{21}(v)+g(u,v)\big(t_{11}(v)t_{22}(u)-t_{11}(u)t_{22}(v)\big).
\end{gather}
The functions $f$ and $g$ are given by (\ref{basicfuncfg}).
They allow us to f\/ind the action on the multiple product of $t_{12}(u)$ that forms a string of length $M$, namely,
\begin{gather*}%\label{comsl21mul}
 t_{11}(u)t_{12}(\bar v)=f(\bar v,u)t_{12}(\bar v)t_{11}(u)+\sum_{i=1}^M g(u,v_i)f(\bar v_i,v_i)t_{12}(u)t_{12}(\bar v_i)t_{11}(v_i),\\
%\label{comsl22}
 t_{22}(u)t_{12}(\bar v)=f(u,\bar v)t_{12}(\bar v)t_{22}(u)+\sum_{i=1}^M g(v_i,u)f(v_i,\bar v_i)t_{12}(u)t_{12}(\bar v_i)t_{22}(v_i),\\
t_{21}(u)t_{12}(\bar v)=t_{12}(\bar v)t_{21}(u)\\
\hphantom{t_{21}(u)t_{12}(\bar v)=}{}
+\sum_{i=1}^{M}f(u_i,\bar u_i)\big(g(u,u_i)t_{22}(u_i)t_{12}(\bar u_i)t_{11}(u)+g(x_i,y)t_{22}(u)t_{12}(\bar u_i)t_{11}(u_i)\big).
\end{gather*}

\section{Projection of the Bethe vector} \label{appB}
The Bethe vector in terms of the operator $t_{12}(u)$ is given by
\begin{gather*}
\nu_{12}(\bar u)\rvec= \mu^N\sum_{i=0}^{N} \sum_{\bar u \to \{\bar u_{\so},\bar u_{\st}\}}\left(\frac{\rho}{\km} \right)^{N-i} \, W^{N}_{i}(\bar u_{\so}|\bar u_{\st})t_{12}(\bar u_{\st})\rvec,
\end{gather*}
with  $\#\bar u_{\st}=i$, $\#\bar u_{\so}=N-i$ a partition of the set $\bar u$. The sum is over all ordered partitions, denoted $\bar u \to \{\bar u_{\so},\bar u_{\st}\}$. The coef\/f\/icient is given by
\begin{gather*}
W^{N}_{i}(u_{1}, \dots , u_{N-i}|u_{N-i+1},\dots , u_{N})=\operatorname{Sym}^{N-i}_{u_{1}, \dots , u_{N-i}}\left(\prod_{j=1}^{N-i}W^{N+1-j}_{N-j}(u_{j}| u_{j+1}, \dots ,u_{N})\right),
\end{gather*}
where
\begin{gather*}
W^{i}_{i-1}(u_{1}| u_{2}, \dots ,u_{i})=f(\bar u_1,u_1)\lambda_1(u_1)+f(u_1,\bar u_1)\lambda_2(u_1)=\Lambda^{i-1}_d(u_1,\bar u_1,1,1)
\end{gather*}
and
\begin{gather*}
\mbox{Sym}^{M}_{\bar u}\big(F(\bar u)\big)=\frac{1}{M!}\sum_{\sigma \in S_M}F(\bar u^\sigma),
\end{gather*}
with $\bar u^\sigma=\{u_{\sigma(1)},\dots,u_{\sigma(M)}\}$ an element of the permutation group~$S_M$. The dual Bethe vector in terms of the operator $t_{21}(u)$ is given by
\begin{gather*}
\lvec\nu_{12}(\bar u)= \mu^N\sum_{i=0}^{N} \sum_{\bar u \to \{\bar u_{\so},\bar u_{\st}\}} \left(\frac{\rho}{\kp} \right)^{N-i} W^{N}_{i}(\bar u_{\so}|\bar u_{\st})\lvec t_{21}(\bar u_{\st}).
\end{gather*}

\subsection*{Acknowledgements} We thank J.~Avan, N.~Grosjean and
R~ Nepomechie for discussions.
R.A.P.\ would like to thanks the hospitality of the Laboratoire de Physique Th\'{e}orique et Mod\'{e}lisation at
the Universit\'e de Cergy-Pontoise where a part of this work was done. S.B.\ is supported by the Universit\'e de Cergy-Pontoise post doctoral fellowship.  R.A.P.\ is supported by Sao Paulo Research Foundation (FAPESP), grants \# 2014/00453-8 and \# 2014/20364-0. We also thank the referees for their constructive remarks.

\pdfbookmark[1]{References}{ref}
\LastPageEnding

\end{document}